\documentclass[%
 reprint,
 amsmath,amssymb,
 aps,
]{revtex4-2}
\usepackage{braket}
\usepackage{here}
\usepackage[dvipdfmx]{graphicx}
\usepackage{dcolumn}
\usepackage{bm}
\usepackage{newtxtext,newtxmath}
\usepackage{hyperref}
\hypersetup{
setpagesize=false,
bookmarksnumbered=false,%
bookmarksopen=false,%
colorlinks=true,%
linkcolor=blue,
citecolor=blue,
urlcolor=blue,
}
\usepackage[normalem]{ulem}
\usepackage{color}

\newcommand{\bs}{\boldsymbol}

\newcommand{\Add}[1]{{#1}}

\begin{document}

\title{Flow patterns and defect dynamics of active nematics under an electric field}

\author{Yutaka Kinoshita}
\author{Nariya Uchida}%
 \email{uchida@cmpt.phys.tohoku.ac.jp}
\affiliation{%
 Department of Physics, Tohoku University, Sendai 980-8578, Japan
}%

\date{\today}

\begin{abstract}
The effects of an electric field on the flow patterns and defect dynamics of 
two-dimensional active nematics are numerically investigated.
We found that field-induced director reorientation 
causes anisotropic active turbulence
characterized by enhanced flow perpendicular to the electric field. 
The average flow speed  and its anisotropy are maximized at an intermediate field strength.
Topological defects in the anisotropic active turbulence are localized and 
show characteristic dynamics \Add{with simultaneous creation of} two pairs of defects.
\Add{A laning state characterized by stripe domains with alternating flow directions} 
is found at a larger field strength near the transition 
to the uniformly aligned state.
We obtained periodic oscillations between the laning state and active turbulence, 
which resembles an experimental observation
of active nematics subject to anisotropic friction.
\end{abstract}


\maketitle


\section{\label{sec::intro}Introduction}

Active hydrodynamics of self-driven particles have attracted much attention
in the last two decades~\cite{simha2002hydrodynamic,
lauga2009hydrodynamics,marchetti2013hydrodynamics}.
Hydrodynamic instabilities lead to mesoscale active 
turbulence~\cite{alert2022active} as found in 
bacterial suspensions~\cite{sokolov2007concentration,wolgemuth2008collective,wensink2012meso} and
microtubule-motor suspensions~\cite{sanchez2012spontaneous,henkin2014tunable}.
Collective hydrodynamics of active elements that have elongated shapes
and apolar interactions 
is described by active nematics~\cite{doostmohammadi2018active}.
While the dynamics of passive nematics are controlled by
external fields and stresses, 
active nematics are internally stirred by active stresses,
which produce chaotic flow patterns containing a number of vortices and topological defects.
Emergence and development of active turbulence are theoretically 
described as follows~\cite{thampi2014instabilities};
(i) aligned regions deform by hydrodynamic instability and create walls, 
which store large elastic energy due to bend deformation;
(ii) the walls collapse by forming pairs of $\pm 1/2$ topological defects;
(iii) the $+1/2$ defects self-propel while the $-1/2$ defects 
are advected by flow;
(iv) a pair of opposite-charge defects attract each other and annihilate by collision.
The dynamical states of microtubule-motor suspensions are controlled by 
ATP concentration~\cite{henkin2014tunable,lemma2019statistical},
confinement~\cite{doostmohammadi2017onset,shendruk2017dancing,suzuki2017spatial,norton2018insensitivity,coelho2019active,
rorai2021active}, and friction~\cite{guillamat2016control,guillamat2016probing,guillamat2017taming,
thampi2014active,srivastava2016negative,thijssen2020active}.
The effects of anisotropic friction induced by a magnetic field are explored in
Ref.~\cite{guillamat2016control}.
\Add{A laning state, in which flow with opposite directions coexist and form stripe domains,
is observed with its flow directions perpendicular to the field.
}
The experimetal study also finds oscillations of the average flow speed and defect density due to 
an instability of the laning state.
Laning states are also obtained in numerical simulations with isotropic~\cite{thampi2014active,srivastava2016negative} 
or anisotropic~\cite{thijssen2020active} friction.

On the other hand, the effects of an electric field on active nematics 
are relatively unexplored.
\Add{An individual microtubule is reoriented parallel to 
a DC~\cite{ramalho2007microtubule,van2006molecular,kim2007active,isozaki2015control} 
and an AC~\cite{minoura2006dielectric} electric field, 
which is explained by the negative net charge and dielectric polarizability of microtubules, respectively. }
However, an experimental study on
collective motion of microtubule-kinesin suspensions under an electric field is lacking. 
Theoretically, a model of an active pump using a Frederiks twisted cell is proposed~\cite{green2017geometry}.
A three-dimensional simulation of active nematics under an electric field
finds a smooth decrease of the defect density with increased field strength 
and a transition to a uniformly aligned state~\cite{krajnik2020spectral}.

In the present paper, we numerically study the effects of an electric field 
on two-dimensional active nematics, focusing on the flow pattern and defect dynamics.
\Add{We assume alignment of the nematic director along the electric field
due to dielectric anisotropy, neglecting electrophoretic or electroosmotic effects.
}
We find that the flow is enhanced in the directions perpendicular to the electric field.
The average flow speed and its anisotropy change non-monotonically with the field strength. 
We obtained a laning state at a larger field strength near the transition to a uniformly aligned state,
and periodic oscillations between the laning state and anisotropic turbulence.
Topological defects are localized in a single vortex region and
show simultaneous creation and recombination of two pairs of defects.

\section{\label{sec::model}Model}

\subsection{\label{subsec::squations}Equations}

We consider a uniaxial active nematic liquid crystal in two dimensions.
The orientational order is described by the symmetric and traceless 
tensor $Q_{ij}=S\left(n_i n_j - \frac12 \delta_{ij} \right)$, 
where $S$ is the scalar order parameter 
and $\bs{n}=\left(\cos\theta,\sin\theta\right)$ is the director.
The dynamical equations are written
in the dimensionless form~\cite{doostmohammadi2018active} 
\begin{align}
\left(\partial_t+\boldsymbol{v}\cdot\boldsymbol{\nabla}\right)\bs{v}
&=
\frac{1}{\mathrm{Re}}\Add{\nabla^2} \bs{v}-\bs{\nabla}p+\bs{\nabla}\cdot\bs{\upsigma},
\label{eq:dvdt}
\\
\bs{\nabla}\cdot\bs{v}
&=
0,
\label{eq:divv}
\\ 
\left(\partial_t+\boldsymbol{v}\cdot\boldsymbol{\nabla}\right)\mathbf{Q}
&=
\lambda S\mathbf{u}+\mathbf{Q}\cdot\bs{\upomega}-\bs{\upomega}\cdot\mathbf{Q}+\gamma^{-1}\mathbf{H}.
\label{eq:dQdt}
\end{align}

Here, $\bs{v}$ is the normalized flow velocity, $p$ is the pressure and 
$\bs{\upsigma}$ is the stress tensor. 
The flow properties are characterized by
the Reynolds number  $\mathrm{Re}$, 
the flow alignment parameter $\lambda$, 
and the rotational viscosity $\gamma$,
and 
$u_{ij}=(\partial_i v_j+\partial_j v_i)/2$ and $\upomega_{ij}=(\partial_i v_j-\partial_j v_i)/2$ 
are the symmetric and antisymmetric parts of velocity gradient tensor, respectively.
Hereafter we call $\omega\coloneq\omega_{xy}$ the vorticity.
\Add{We assume $0<\lambda <1$; a positive value of $\lambda$
corresponds to rod-like or filamentous material and 
$|\lambda|<1$ means that there is no stable director orientation
in a uniform shear flow 
(flow-tumbling regime)~\cite{de1993physics, edwards2009spontaneous}.
}
The molecular field $H_{ij}$
\Add{
is defined as the symmetric and traceless part of $-\delta F/\delta Q_{ij}$, 
and}
is obtained from the Landau-de Gennes free energy\Add{~\cite{de1993physics}}
\begin{align}
F&=\int f \mathrm{d}^2 r,
\label{eq:F}
\\
f&= 
\frac{A}{2}\mathrm{Tr} \, \mathbf{Q}^2
+ 
\frac{C}{4}\left(\mathrm{Tr} \, \mathbf{Q}^2\right)^2
+ 
\frac{K}{2}\left(\bs{\nabla}\mathbf{Q}\right)^2
-
\frac{\varepsilon_a}{2}\bs{E}\cdot\mathbf{Q}\cdot\bs{E}.
\label{eq:f}
\end{align}
The first two terms of the free energy density control the magnitude of the scalar order parameter $S$, 
the third term is the Frank elastic energy under the one-constant approximation,
and the fourth term describes the coupling to the electric field.
\Add{
Here, we dropped the term proportional to $\mathrm{Tr}\, \mathbf{Q}^3$ 
contained in the standard expression of 
the Landau de Gennes free energy, because it identically vanishes 
for the two-dimensional nematic order parameter.
}
We assume positive dielectric anisotropy ($\varepsilon_a>0$) 
so that the director tends to align parallel to the electric field,
\Add{mimicking the response of an individual microtubule~\cite{kim2007active,minoura2006dielectric}.}
\Add{
The molecular field is obtained as
\begin{align}
H_{xx} &= - \left(2 A + C S^2 \right) Q_{xx} + 2 K \nabla^2 Q_{xx} 
+ \frac{\varepsilon_a}{2} \left(E_x^2 - E_y^2 \right).
\label{eq:Hxx}
\\
H_{xy} &= - \left(2 A + C S^2 \right) Q_{xy} + 2 K \nabla^2 Q_{xy} 
+ \varepsilon_a E_x E_y.
\label{eq:Hxy}
\end{align}
}
The stress tensor is the sum of the passive stress 
\begin{equation}    
\bs{\upsigma}^\mathrm{e}=
-\lambda S \mathbf{H}+\mathbf{Q}\cdot\mathbf{H}-\mathbf{H}\cdot\mathbf{Q},
\label{eq:sigmae}
\end{equation}
and the active stress
\begin{equation}    
\bs{\upsigma}^\mathrm{a}=-\alpha\mathbf{Q}.
\label{eq:sigmaa}
\end{equation}
We assume that the activity parameter $\alpha$ 
is positive, which corresponds to an extensile system.

\subsection{\label{sec::linstability}Linear stability analysis}

In the absence of an electric field,  the model exhibits macroscopically isotropic
active turbulence. 
\Add{Because we assumed positive dielectric anisotropy,}
the electric field is expected to align the director 
and stabilize a uniform ordered state if the field is sufficiently strong.
We conducted a linear stability analysis of the uniformly aligned state
to find the threshold electric field and analyzed the onset of instability.
We assume a uniform electric field along the $x$-axis ($\bm{E} = E \bm{e}_x$).
We also assume that the system 
is far from the nematic-isotropic transition point and 
that the scalar order parameter is given by
the equilibrium value $S_0$, which satisfies
\begin{equation}
 -S_0\left(A+\frac{C}{2}S_0^2\right)+\frac{\varepsilon_a}{2}E^2=0.
 \label{eq:S0} 
\end{equation}

In the unperturbed quiescent state, 
the director is aligned along the $x$-axis so that   
\begin{equation}
    \bs{v}_0 =\bs{0},
    \quad
    Q_{xx,0} =\frac{S_0}{2},
    \quad 
    Q_{xy,0}=0, 
    \label{eq:Q0v0}
\end{equation}
where the subscript 0 means the unperturbed state.
We consider perturbations to
the flow velocity $\delta\bs{v}$
and
the director angle $\delta\theta$,
which gives the variations of
the order parameter 
\begin{align}
\delta Q_{xx} &=\frac{S_0}{2} [\cos (2\delta \theta) -1]  \simeq 0,
\label{eq:deltaQxx}
\\
\delta Q_{xy} &=\frac{S_0}{2} \sin (2\delta \theta) \simeq S_0 \delta \theta.
\label{eq:deltaQxy}
\end{align}

Assuming the time dependence proportional to $\mathrm{e}^{-\mathrm{i}\omega t}$,
where $\omega$ is the complex frequency, we expand the director angle and flow velocity in the Fourier modes
as
\begin{align}
\theta(\bm{r}, t) 
&= 
\frac{1}{(2\pi)^2}\int \mathrm{d}^2 k \, 
\delta \theta^{\bm{k}} e^{\mathrm{i} \left(\bm{k}\cdot\bm{r}-\omega t \right)},
\label{eq:thetak}
\\
\bm{v}(\bm{r}, t) 
&= 
\frac{1}{(2\pi)^2}\int \mathrm{d}^2 k  \,
\delta \bm{v}^{\bm{k}} e^{\mathrm{i} \left(\bm{k}\cdot\bm{r}-\omega t \right)},
\label{eq:vk}
\end{align}
where 
$\bm{k}= (k_x, k_y) = \left(k\cos\phi,k\sin\phi\right)$ is the wavevector.
Because $\theta = 0$ and $\bm{v}=0$ in the unperturbed state, 
we consider only the modes with nonzero wavevectors. 
\Add{
Similarly, we expand $Q_{ij}$, $H_{ij}$, $\sigma_{ij}$ and $p$ in the Fourier space
and then to the first order of $\delta \theta^{\bm{k}}$ and $\delta \bs{v}^{\bm{k}}$.
From Eqs.(\ref{eq:deltaQxx},\ref{eq:deltaQxy}), we have 
$\delta Q_{xx}^{\bm{k}} \simeq 0$ and 
$\delta Q_{xy}^{\bm{k}} \simeq S_0 \delta \theta^{\bm{k}}$.
The molecular field vanishes in the unperturbed state: $H_{ij,0} =0$.
Its variations are  read from Eqs.(\ref{eq:Hxx},\ref{eq:Hxy}) as
$\delta H_{xx}^{\bm{k}} \propto \delta Q_{xx}^{\bm{k}} \simeq 0$ 
and 
\begin{equation}
\delta H_{xy}^{\bm{k}} \simeq 
\left[-\left( 2 A + C S_0^2 \right) - K k^2\right] \delta Q_{xy}^{\bm{k}} 
\simeq - D \delta \theta^{\bm{k}},
\label{eq:deltaHxy}
\end{equation}
where we introduced the abbreviation 
\begin{equation}
D = D(k) \equiv \varepsilon_a E^2 + 2 K S_0 k^2 
\label{eq:D}
\end{equation}
and used Eq.(\ref{eq:S0}).
For the stress tensor, 
we readily obtain 
$\delta \sigma_{xx} = -\delta \sigma_{yy} 
\simeq - \lambda S_0 \delta H_{xx} - \alpha \delta Q_{xx}\simeq 0$,
\begin{align}
\delta \sigma_{xy}^{\bm{k}} 
& \simeq (1 -\lambda) S_0 \delta H_{xy}^{\bm{k}} - \alpha \delta Q_{xy}^{\bm{k}}
\nonumber \\
& \simeq \left[ (\lambda-1) D - \alpha \right] S_0 \delta \theta^{\bm{k}},
\label{eq:sigma_xy}
\end{align}
and 
\begin{align}
\delta \sigma_{yx}^{\bm{k}} 
& \simeq -(1 + \lambda) S_0 \delta H_{xy}^{\bm{k}} - \alpha \delta Q_{xy}^{\bm{k}}
\nonumber \\
& \simeq \left[ (\lambda+1) D - \alpha \right] S_0 \delta \theta^{\bm{k}}.
\label{eq:sigma_yx}
\end{align}
The equation of motion (\ref{eq:dvdt}) is linearized as
\begin{align}
- \mathrm{i} \omega \delta v_{i}^{\bm{k}} 
= 
-\frac{k^2}{\rm Re} \delta v_{i}^{\bm{k}}
- \mathrm{i} k_{i} \delta p^{\bm{k}}
+ \mathrm{i} k_{j} \delta \upsigma_{ij}^{\bm{k}}.
\label{eq:ddeltavdt}
\end{align} 
Summation over repeated indices ($i,j = x,y$) is assumed here and hereafter.
The pressure in (\ref{eq:ddeltavdt}) is determined by
the incompressibility condition (\ref{eq:divv}), 
or equivalently $\bm{k} \cdot \delta \bm{v}^{\bm{k}} = 0$,
as
\begin{align}
\delta p^{\bm{k}} = 
\hat{k}_i \hat{k}_j  \delta \sigma_{ij}^{\bm{k}},
\label{eq:p}
\end{align} 
where we introduced 
$\hat{\bm{k}} = \bm{k}/|\bm{k}| = (\hat{k}_x, \hat{k}_y) = (\cos \phi, \sin \phi)$.
The incompressibility condition allows us to 
express the velocity by the perpendicular component 
\begin{equation}
\delta v_\perp^{\bs{k}} 
= -\hat{k}_y \delta v_x^{\bs{k}} 
+  \hat{k}_x \delta v_y^{\bs{k}}
\label{eq:deltavperp}
\end{equation}
as
\begin{equation}
\delta \bm{v}^{\bs{k}} 
= (\delta v_x^{\bs{k}}, \delta v_y^{\bs{k}}) 
= \delta v_\perp^{\bs{k}} (-\hat{k}_y, \hat{k}_x).
\label{eq:deltav}
\end{equation}
Rewriting (\ref{eq:ddeltavdt})  in terms of $\delta v_\perp^{\bs{k}}$ 
and substituting (\ref{eq:sigma_xy},\ref{eq:sigma_yx}),
we get 
\begin{align}
- \mathrm{i} \omega \delta v_{\perp}^{\bm{k}} 
&= 
-\frac{k^2}{\rm Re} \delta v_{\perp}^{\bm{k}}
- \mathrm{i}k \left( 
  \hat{k}_y^2 \delta \sigma_{xy}^{\bm{k}} 
- \hat{k}_x^2 \delta \sigma_{yx}^{\bm{k}} \right)
\nonumber \\
&=
-\frac{k^2}{\rm Re} \delta v_{\perp}^{\bm{k}}
- \mathrm{i}k \left[ 
  \left( \hat{k}_y^2 - \hat{k}_x^2 \right) \left( \lambda D - \alpha \right) -D 
  \right]
  S_0 \delta \theta^{\bm{k}}.
\label{eq:dvperpdt} 
\end{align} 
On the other hand, the dynamical equation (\ref{eq:dQdt}) 
for the order parameter 
is linearized as
\begin{align}
- \mathrm{i} \omega S_0 \delta \theta^{\bm{k}} 
&= S_0
\left(
\frac{\lambda+1}{2} \mathrm{i} k_x \delta v_y^{\bm{k}} 
+
\frac{\lambda-1}{2} \mathrm{i} k_y \delta v_x^{\bm{k}} 
\right)
- \frac{1}{\gamma} D \delta \theta^{\bm{k}} 
\nonumber \\
&=
\mathrm{i} k S_0 
\left(
\frac{\lambda+1}{2} \hat{k}_x^2 
-
\frac{\lambda-1}{2} \hat{k}_y^2 
\right)
\delta v_\perp^{\bm{k}} 
- \frac{1}{\gamma} D \delta \theta^{\bm{k}} 
\label{eq:dthetadt}
\end{align}
with the aid of Eqs.(\ref{eq:deltaQxy},\ref{eq:deltaHxy},\ref{eq:D},\ref{eq:deltav}).
Eqs.(\ref{eq:dvperpdt},\ref{eq:dthetadt}) are written in the matrix form
\begin{equation}
        -\mathrm{i}\omega
        \begin{pmatrix}
            \delta v_\perp^{\bs{k}}
            \\
            \delta \theta^{\bs{k}} 
        \end{pmatrix}
    = {\cal M}
        \begin{pmatrix}
            \delta v_\perp^{\bs{k}}
            \\
            \delta \theta^{\bs{k}} 
        \end{pmatrix}
\label{eq:matrix}
\end{equation}
with 
\begin{equation}
{\cal M} =
        \begin{pmatrix}
        - \frac{k^2}{\mathrm{Re}}
       &
      \mathrm{i} k S_0 \left[  \left(\lambda \cos 2\phi +  1 \right) D - \alpha \cos2\phi
       \right] 
\vspace{3mm}
\\
       \mathrm{i} k \frac{\lambda \cos 2\phi +  1}{2}
       &  
       - \frac{D}{\gamma S_0} 
    \end{pmatrix}.
\label{eq:matrixM}
\end{equation}    
A straightforward calculation gives
the eigenvalues $-\mathrm{i}\omega_{\pm}^{\bm{k}}$  of ${\cal M}$,
and the stability condition is obtained 
as $\mathrm{Im} \, \omega_{\pm}^{\bm{k}} \le 0$  for any $\bm{k}$;
see Appendix \ref{sec:eigenvalues} for details. 
We identified the most unstable mode that maximizes
the linear growth rate $\mathrm{Im} \, \omega^{\bm{k},j}$ ($j=1,2$)
for given $\alpha$ and $E$.
For a given angle $\phi$ of the wavevector, 
the mode in the long-wavelength limit ($k\to 0$) is most unstable, 
which is trivial since Frank elasticity suppresses deformation.
For a given wavenumber $k$, 
the mode with the $\phi=0$ is most unstable 
in the range $0 < \lambda < 1$ assumed in this paper.
This is in accordance with the general observation 
on extensile (or "pusher"-type) active matter 
that the active stress induces bend deformation of the axis of alignment~\cite{simha2002hydrodynamic,baskaran2009statistical}, 
which is interpreted as a buckling instability of the filaments~\cite{vliegenthart2020filamentous}.
The stability condition is given by
\begin{align}
\alpha < \alpha_c = \left(\lambda + 1 + \frac{B}{\lambda+1} \right) \varepsilon_a E^2,
\quad 
B = \frac{2}{\gamma S_0^2 \mathrm{Re}}. 
\label{eq:alphac}
\end{align}    
Note that $B$ weakly depends on $E$ via the cubic equation (\ref{eq:S0}) for $S_0$,
and the analytical expression of $\alpha_c = \alpha_c(E)$ becomes voluminous.
For numerical analysis, we choose the parameter values 
\begin{align}
A&=-0.15, \,
C=0.4, \,
K=0.5, \,
\varepsilon_a=1,
\label{eq:parameter1}
\\
\lambda&=0.1, \, \mathrm{Re}=0.1,\, \gamma=10.
\label{eq:parameter2}
\end{align}
We varied the field strength in the range $0 \le E \le 0.3$.
The stability threshold is given by 
$\alpha_c = 0.1$ for $E\simeq 0.175$ and
$\alpha_c=0.2$ for $E\simeq 0.255$, for example. 
}

\begin{figure}[thb]
\includegraphics{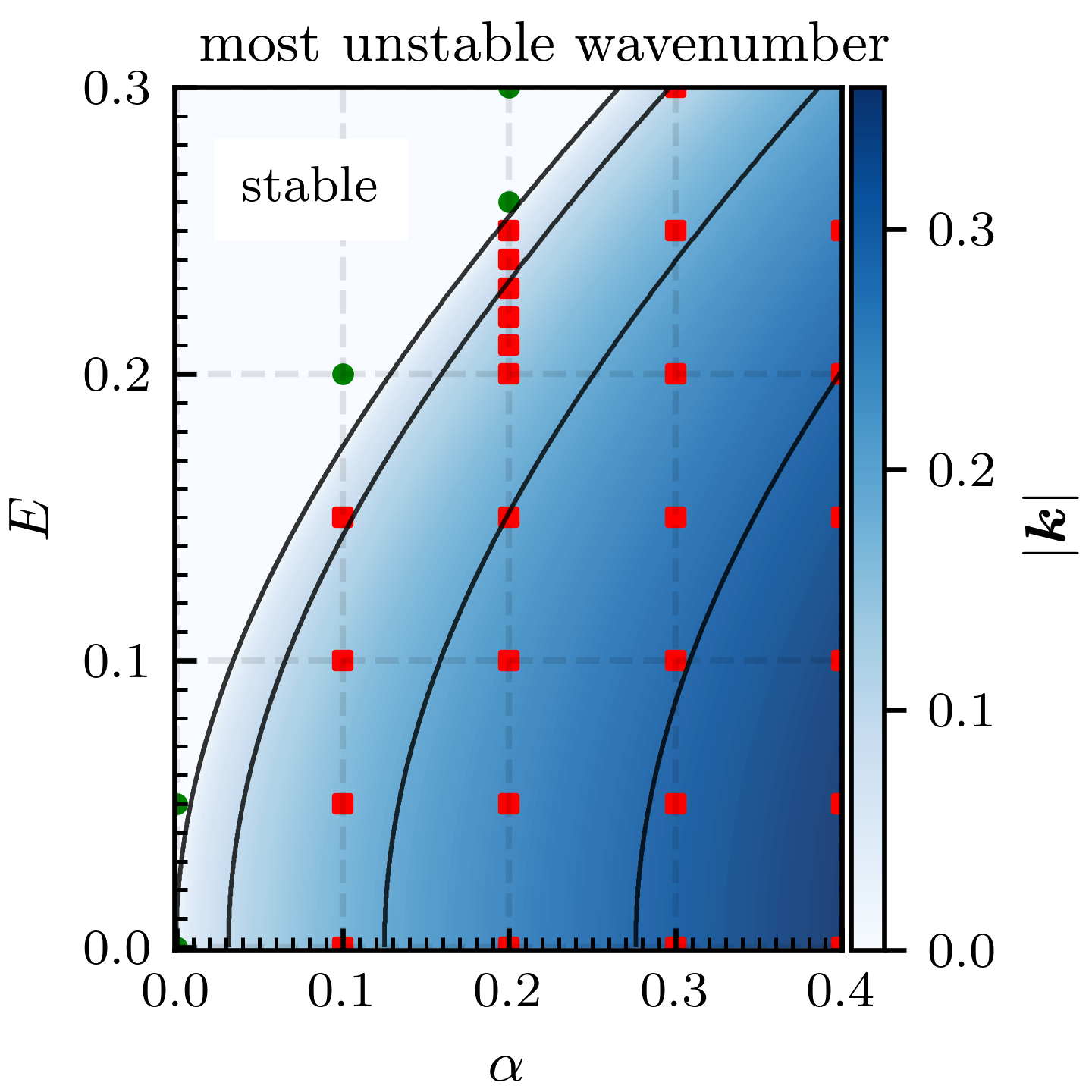}
\caption{\label{fig:linstability} 
Results of the linear stability analysis for a uniformly aligned state.
Contour lines for the most unstable wavenumber  
\Add{
$|\bs{k}| = 0, 0.1, 0.2, 0.3$ 
}
are shown in solid lines.
The stability threshold coincides with the contour line for $k=0$.
\Add{
For a given value of $|\bs{k}|$, 
the 
}
direction of the most unstable wavevector is parallel 
to the unperturbed director.
The result of numerical simulations is shown by circles (squares)
where the uniformly aligned state is stable (unstable).
}
\end{figure}

In Fig.\ref{fig:linstability}, we plot the magnitude of the most unstable wavenumber 
as a function of $\alpha$ and $E$. 
The region $|\bm{k}|=0$ is identical with the linearly stable region.
The contour line for $|\bm{k}|=0$ means the stability threshold
and starts from the origin $(\alpha, E) = (0,0)$.
The most unstable wavenumber increases as $\alpha$ increases or $E$ decreases.

\section{\label{sec::results}Numerical Results}

\subsection{\label{subsec::method}Method and parameters}

We numerically solved the equations Eqs.(\ref{eq:dvdt},\ref{eq:divv},\ref{eq:dQdt})
on a square lattice with the fourth-order Runge-Kutta method.
The incompressibility condition is handled by the simplified MAC method 
on a staggered lattice~\cite{amsden1970simplified}.
The main sublattice is used for $\mathbf{Q}(\bm{r},t)$ and the auxiliary field variables 
$p$, $\boldsymbol{\upsigma}$, $\mathbf{u}$, $\boldsymbol{\upomega}$ 
and $\mathbf{H}$,
and the other two are assigned to $v_x(\bm{r}, t)$ and $v_y(\bm{r},t)$.
The calculation is performed on a $N_x \times N_y$ lattice 
with the grid size $\Delta x = \Delta y = 2$ 
and the periodic boundary conditions,
with the step time increment $\Delta t=0.01$.
We used the Fast Fourier Transform to solve the Laplace equation
for the pressure at each time step.
The parameter values used in the simulation are given 
\Add{
in Eqs.(\ref{eq:parameter1},\ref{eq:parameter2}).
}
The system size is fixed to 
$N_x=N_y=128$ so that $L=N_x \Delta x = N_y \Delta y = 256$.
In a equilibrium state that minimizes the Landau-de Gennes free energy
with $E=0$,
the scalar order parameter is given by $S_0= \sqrt{2|A|/C} \simeq 0.87$
and the defect core radius is 
\Add{$\xi = \sqrt{K/|A|} \simeq 1.83$}.
We varied the activity parameter $\alpha$ and the electric field strength $E$.
For the initial conditions,
we set the velocity to zero and assumed small random fluctuations
around zero for $\mathbf{Q}(\bm{r},0)$, 
assuming a quench from the isotropic quiescent state.
We observed the total kinetic energy as a function of time
to confirm that the system reached dynamical steady states,
typically by $t = 10000$ for active turbulence states.
We calculated statistical data over the time window $t_0 < t < t_0 + 20000$
after the system reached the dynamical steady state ($t_0$ is varied depending on the parameter).

\subsection{\label{subsec:dynamical} Spatiotemporal patterns and flow anisotropy}

\begin{figure}[htb]
\centering
\includegraphics[width=160mm,pagebox=cropbox]{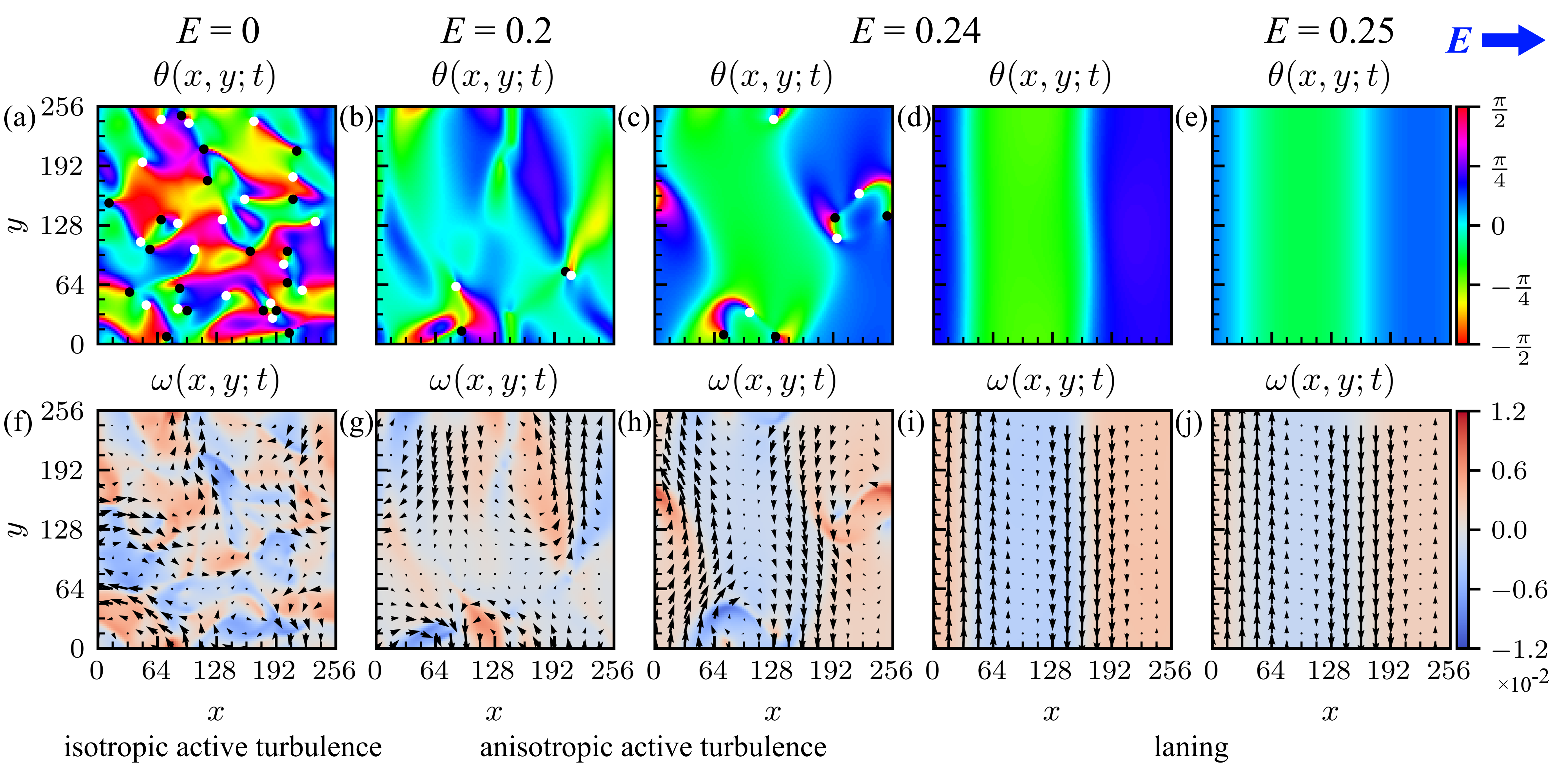}
\caption{\label{fig:snapshot} 
Snapshots 
of the director angle $\theta(x,y)$ (top row)
and the vorticity $\omega(x,y)$ with the velocity field shown by arrows
(bottom row)
in the dynamical steady states for $\alpha=0.2$. 
(a),(f) Isotropic active turbulence ($E=0$).
(b),(g) Anisotropic active turbulence ($E=0.2$).
We find a periodic switching between 
(c),(h) anisotropic turbulence and 
(d),(i) a laning state   ($E=0.24$).  
(e), (j) A steady laning state ($E=0.25$). 
}
\end{figure}

In Fig.\ref{fig:snapshot}, we show the snapshots of the director angle $\theta(\bm{r}, t)$ 
and the vorticity $\omega(\bm{r},t)$ in the dynamical steady states
for $\alpha =0.2$.
For $E=0$, we obtain active turbulence containing topological defects 
and vortices that are macroscopicaly isotropic [Fig.\ref{fig:snapshot}(a)(f)].
For $E=0.2$, 
the director is tilted toward the electric field and forms 
anisotropic active turbulence
with fewer defects and vortices compared to the zero-field case
[Fig.\ref{fig:snapshot}(b)(g)].
For $E=0.24$, we observe a periodical switching between 
the anisotropic turbulence 
[Fig.\ref{fig:snapshot}(c)(h)]
and a laning state with bidirectional flow 
[Fig.\ref{fig:snapshot}(d)(i)].
A steady laning state is obtained at $E=0.25$
[Fig.\ref{fig:snapshot}(e)(j)].
For $E \ge 0.26$, uniformly aligned state is obtained (not shown).
In Fig.\ref{fig:linstability}, 
we plot the parameter sets where uniformly aligned states are stable (unstable)
by circles (squares), which agree with the result of the linear stability analysis.

\begin{figure}[htb!]
\includegraphics[pagebox=cropbox]{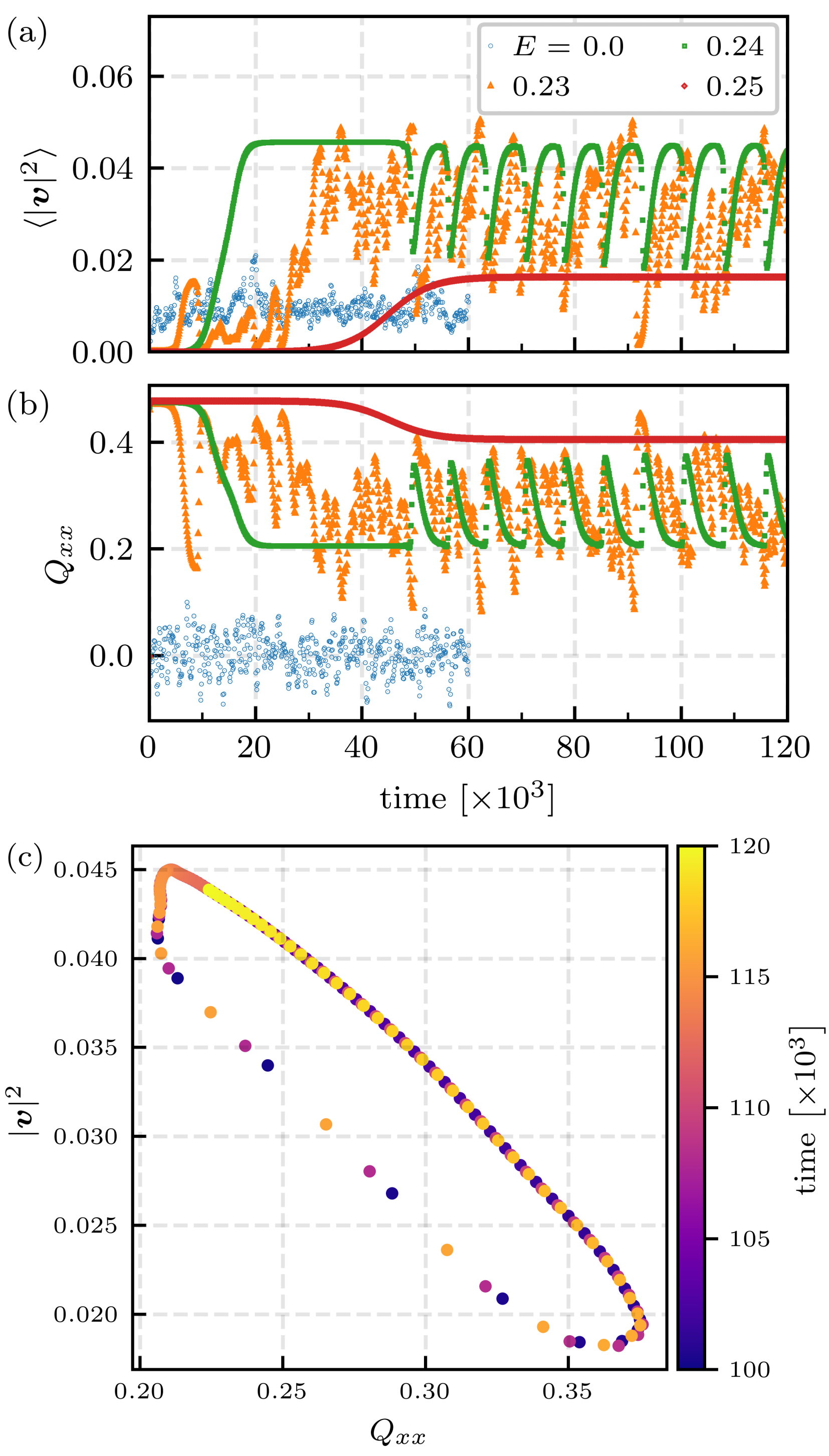}
\caption{\label{fig:v2Q} 
Time development of (a)
the mean square velocity $\langle \boldsymbol{v}^2 \rangle$
and
(b) the order parameter component $\langle Q_{xx} \rangle$.
\Add{(c) Trajectory in the $\langle Q_{xx} \rangle$-$\langle \boldsymbol{v}^2 \rangle$ plane for $E=0.24$.
Time evolution takes place in the anticlockwise direction.
The data are for $\alpha=0.2$.
}
}
\end{figure}

\Add{
In Fig.\ref{fig:v2Q}(a)(b),  we show time evolution of 
the mean square velocity $\langle|\bm{v}^2|\rangle$ 
and 
the order parameter component $\langle Q_{xx}\rangle$, respectively,}
\Add{
for a specific sample (initial condition),
where 
the average is taken over space.}
Compared to the isotropic active turbulence at $E=0$,
the anisotropic turbulence at $E=0.23$ has a longer incubation time 
for the velocity to grow, and has larger mean value and fluctuations 
in the dynamical steady state.
At $E=0.24$, we find periodic oscillations between the anisotropic turbulence
and laning state.
\Add{
The active flow velocity reaches a plateau by $t=50\times 10^3$, 
which corresponds to the laning state. 
Then the velocity rapidly decreases 
as the lane collapses and emits multiple pairs of defects.
The defects slowly annihilate, leaving a lane behind.
This cycle is repeated until the end of the simulation,
with the period $T\simeq 7 \times 10^3$.
}
The laning state for $E=0.25$ has a larger steady state value of
the mean square velocity than the isotropic turbulence.
\Add{
Under the electric field, 
the degree of alignment $\langle Q_{xx} \rangle$ has negative correlation with the velocity,
which is most clearly seen in the oscillations for $E=0.24$.
In Fig.\ref{fig:v2Q}(c), we show time-evolution of the system
in the $\langle Q_{xx} \rangle-\langle|\boldsymbol{v}^2|\rangle$ plane.
The system makes an anticlockwise cycle in the plane, 
with the slow phase characterized by a larger velocity magnitude 
than the rapid phase.
Because the slow phase involves straightening of lanes
it is natural that it generates stronger flow 
than the rapid phase where flow is turbulent and slow.
}
 
\begin{figure}[htb!]
\includegraphics[pagebox=cropbox]{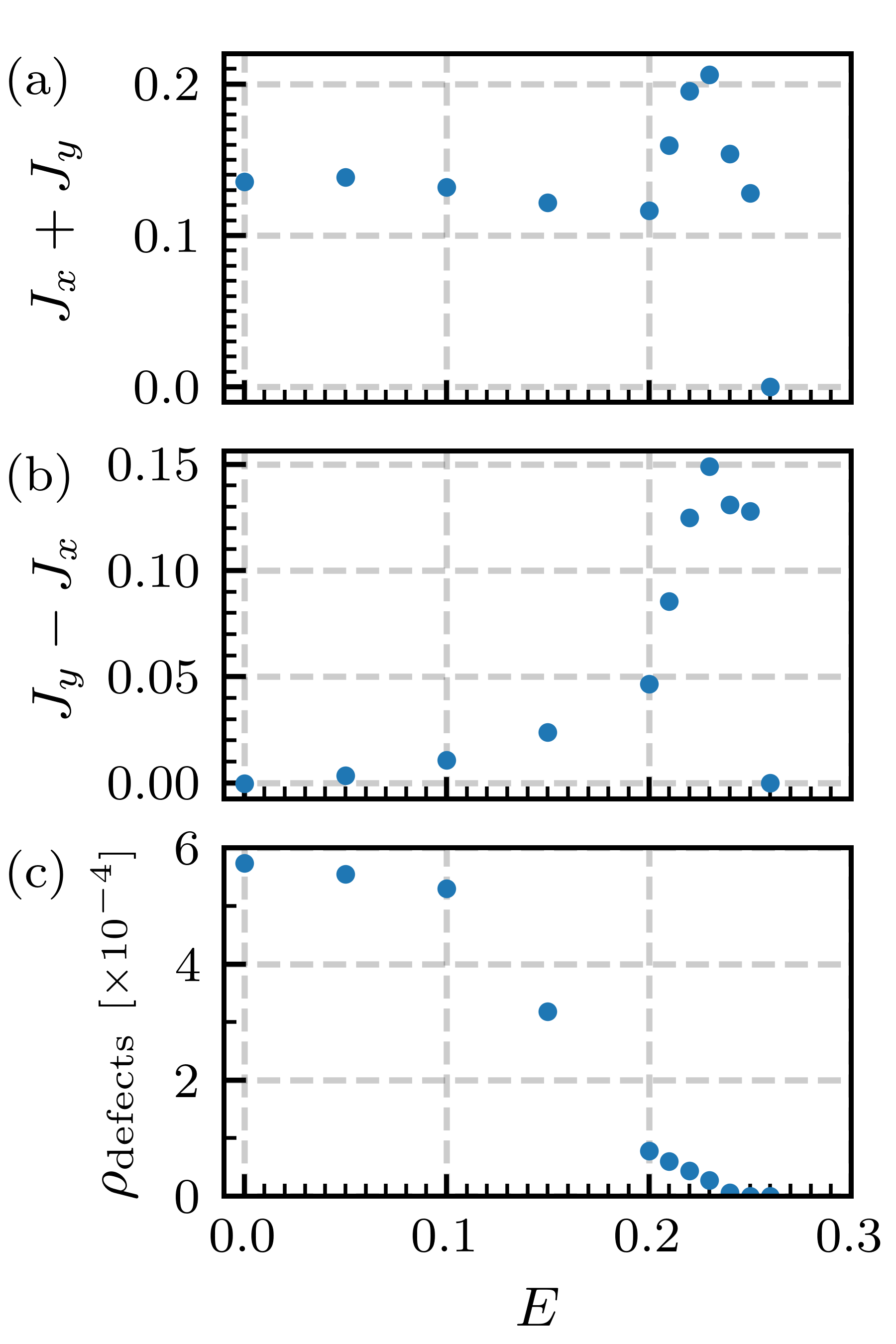}
\caption{\label{fig:flux} 
(a)
Total flux $J_x+J_y$,
(b)
anisotropy of flux $J_x- J_y$,
and
(c)
defect density $\rho_{\rm defect}$ versus the field strength $E$.
The data are for $\alpha=0.2$.
}
\end{figure}

To characterize the orientation of flow, 
we define $J_x = \sqrt{\langle v_x^2 \rangle}$ and $J_y = \sqrt{\langle v_y^2 \rangle}$,
where the average is taken over time and 10 independent samples.
In Fig.\ref{fig:flux}(a)(b), 
we plot the sum $J_x + J_y$ and difference $J_y-J_x$
as functions of $E$.
As $E$ is increased,
the sum $J_x+J_y$ shows a slow decline for $E\le 0.20$ and then
a sharp peak at $E=0.23$, which corresponds 
to the anisotropic active turbulence.
It rapidly decreases to zero as the electric field is further increased.
The flow anisotropy characterized by $J_y-J_x$
increases as $E$ is increased and also has a peak at $E=0.23$.
In Fig.\ref{fig:flux}(c), we plot the number density of topological defects,
which maintains a large value for $E \le 0.1$ and then rapidly decreases 
as a function of $E$.

\subsection{\label{subsec:dist} Distributions and correlation functions}

\begin{figure}[htb!]
\includegraphics[pagebox=cropbox]{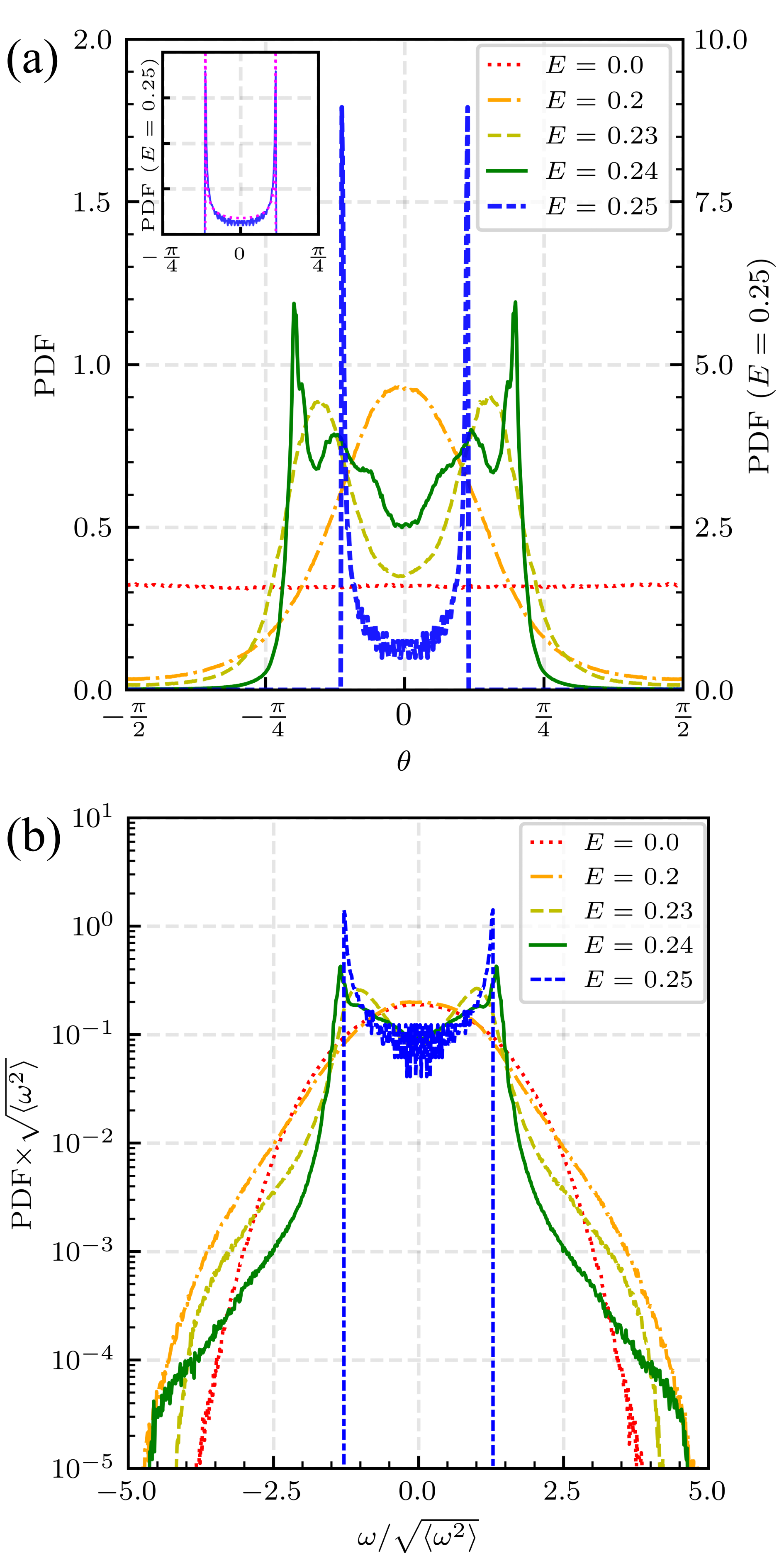}
\caption{\label{fig:distribution} 
Normalized probability distribution functions
of (a) the director angle $\theta$ and 
(b) the vorticity $\omega$, both for $\alpha=0.2$. 
In (a), the data for $E=0.25$ are scaled by the right vertical axis.
\Add{
The inset of (a) shows comparison with the theoretical line for the sinusoidal undulation,
$f(\theta) = 1/[\pi (\theta_0^2-\theta^2)^{1/2}]$ ($|\theta| < \theta_0$, dotted lines).
}
}
\end{figure}

In Fig.\ref{fig:distribution}, we show the probability distribution functions (PDFs)
of the director angle and vorticity for $\alpha=0.2$ 
\Add{calculated  from an ensemble of 10 samples.}
The PDF of the director angle [Fig.\ref{fig:distribution}(a)] 
has a single peak at $\theta=0$ for $0 < E\le 0.20$,
with its height increasing with $E$.
The peak splits into two at $E=0.23$, 
which is interpreted as a precursor of the laning state.
At $E=0.24$, the PDF has four peaks 
\Add{
resulting from coexistence of narrow and wide lanes. 
Depending on the initial condition,
we obtained either narrow lanes with the period equal to $L/2$, 
or wide lanes with the period equal to the system size ($L$).
Each type of lanes sporadically collapse into a defected state and recover,
but do not interchange with each other. 
We find narrow lanes in 4 out of the 10 samples, and wide lanes in 6 samples.
The wide lanes contribute to the two higher peaks of the distribution function
at larger values of $|\theta|$, and narrow lanes contribute to the lower peaks. 
}
The steady laning state at $E=0.25$ has two sharp peaks at 
$|\theta| = \theta_0 \simeq \pi/8$ with no tails.
\Add{
The peak position is closer to $\theta=0$ than the case $E=0.24$,
which means that the undulation is suppressed by the electric field.
}
Note that the PDF for a sinusoidal undulation $\theta(x)=\theta_0 \sin kx$
is proportional to $|dx/d\theta| \propto (\theta_0^2- \theta^2)^{-1/2}$ 
for $|\theta|<\theta_0$ and is zero for $|\theta|>\theta_0$
\Add{(see Appendix \ref{sec:distribution} for derivation).
In the inset of Fig.\ref{fig:distribution}(a),
we see that the formula gives a good approximation of
the PDF for $E=0.25$, with the minimum at $\theta=0$ slightly lower
than the theoretical value. 
}
The PDF for the vorticity [Fig.\ref{fig:distribution}(b)] is nearly Gaussian
for $E=0$, while a small deviation (fat tail) appears at $E=0.20$ and 
then the peak splits into two at $E=0.23$.
The distribution gets narrower tails as $E$ is further increased to $E=0.25$.

\begin{figure}[htb!]
\includegraphics[pagebox=cropbox]{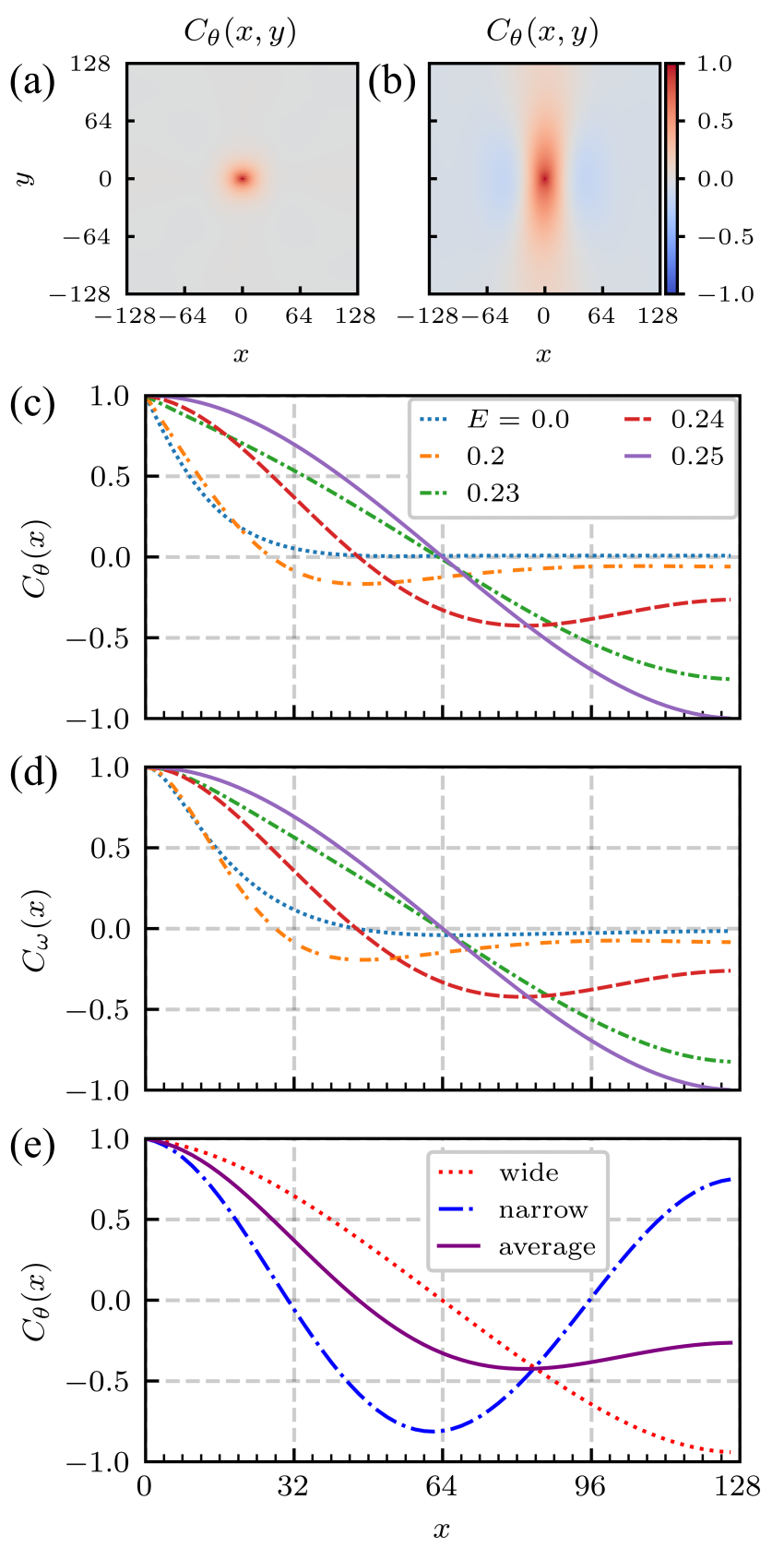}
\caption{\label{fig:correlation-func} 
Spatial correlation functions of the director at (a) $E=0$ and (b) $E=0.2$, both for $\alpha=0.2$.
Spatial correlation functions along the $x$-axis
for 
(c) the director angle and (d) the vorticity.
\Add{(e) $C_{\theta}(x,0)$ for $E=0.24$, 
for narrow lanes (4 samples), wide lanes (6 samples) and the ensemble average.
}
}
\end{figure}

In Fig.\ref{fig:correlation-func}, we plot the spatial correlation functions 
for the director angle and vorticity, which are defined by
\begin{equation}
C_\theta(\bs{r})=\frac{\braket{\theta(\bs{r}+\bs{r'},t)
\theta(\bs{r'},t)}}{\braket{\theta(\bs{r'},t)^2}}
\end{equation}
and
\begin{equation}
C_\omega(\bs{r})=\frac{\braket{\omega(\bs{r}+\bs{r'},t)\omega(\bs{r'},t)}}{\braket{\omega(\bs{r'},t)^2}},
\end{equation}
respectively, where $\braket{\cdots}$ indicates averages over $\bs{r'}$ and $t$,
and 10 independent samples.
The angular correlation function for $E=0$ are isotropic as shown in 
Fig.\ref{fig:correlation-func}(a).
Under an electric field ($E=0.2$), 
the correlation functions develop anisotropic structures
with peaks elongated in the $y$-direction and valleys in the $x$-direction,
as seen in 
Fig.\ref{fig:correlation-func}(b).
\Add{
The correlation profiles along the line $y=0$
are shown in Fig.\ref{fig:correlation-func}(c)(d).
Because 
$C_\theta(x) = C_\theta(x,0)$ and $C_\omega(x) = C_\omega(x,0)$
show very similar behaviors, 
except that the latter shows a slightly non-monotonic decay for $E=0$, 
we focus on the former.
The anisotropic turbulence for $E=0.2$ is characterized by
a shallow minimum at $x\simeq 40$.
For $E=0.23$, 
we observe irregular temporal fluctuations
between a wide lane and a turbulent state, which is reflected 
in the fluctuations of the velocity and order parameter [Fig.~\ref{fig:v2Q}(a)(b)]. 
Since the correlation function for the turbulent state has a shallow minimum. 
the minimum of the time-averaged correlation function
is dominantly determined by that of the wide lane.
This explains the minimum of $C_\theta(x)$ at $x\simeq L$ for $E=0.23$.
For $E=0.24$,
the minimum of $C_{\theta}(x)$ 
is located at a shorter distance ($x\simeq 80$) 
compared to those for $E=0.23$ and $0.25$. 
This is explained by coexistence of narrow and wide lanes 
mentioned in the previous paragraph. 
In Fig.\ref{fig:correlation-func}(c), we show the correlation functions
for the subsets of samples with narrow and wide lanes,
and the ensemble average. 
We see that the average of the correlation functions for narrow and wide lanes
gives a shallow minimum at a distance between $L/2$ and $L$.
Finally, for $E=0.25$, we find only a wide lane, which gives the deep minimum
of $C_\theta(x)$ at $x= L$.
}

\subsection{\label{subsec:defect}Defect dynamics}

\begin{figure*}[t]
\includegraphics[scale=0.95]{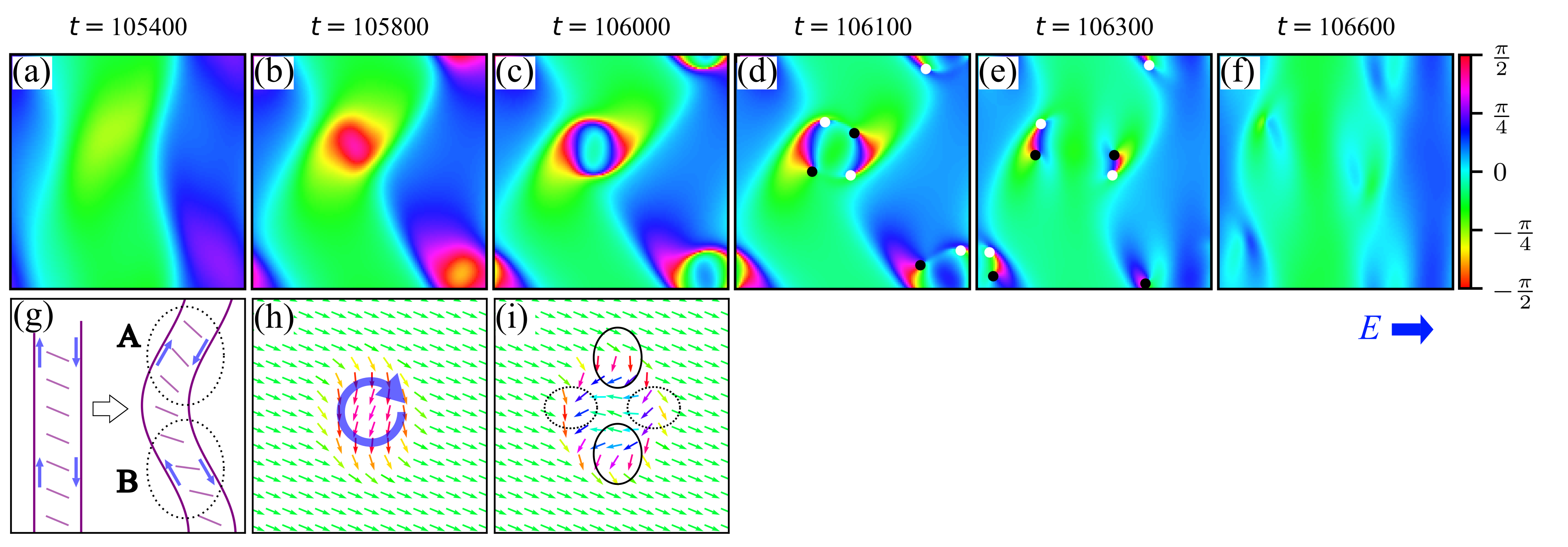}
\caption{\label{fig:dynamics} 
(a)-(f) Snapshots of the defects creation and annihilation process at $\alpha=0.2$ and $E=0.23$;
(a) the director has small tilt angle in the center region;
(b) an island of large tilt angle appears;
(c) further rotation of the director inside the island;
(d) the island breaks up into two crescent-like shapes 
and two pairs of defects are created;
(e) the crescent-shape regions shrink;
(f) defects annihilate by collision.
\Add{
(g)-(i) Schematic pictures of the defect creation process;
(g) a straight lane undergoes a buckling instability and form the oppositely tilted regions A and B.
The active flow along the lane boundary (arrows) generates a torque 
that enhances the clockwise tilt in A, while it suppresses the anti-clockwise tilt in B.
(h) The resultant deformation of the director field induces further rotation in the center region;
(i) Bend deformation in the top and bottom regions (solid lines)
and splay deformation on the left and right hand sides (dotted lines).
The former is transformed  to splay deformation by creating two pairs of defects.
}
}
\end{figure*}

In the anisotropic turbulence state, we observe
characteristic dynamics of topological defects.
A typical time series of defect creation and annihilation is shown in Fig.\ref{fig:dynamics}. 
Starting from a weakly deformed director configuration [Fig.\ref{fig:dynamics}(a)], 
flow-induced instability generates an island of large tilt angle [Fig.\ref{fig:dynamics}(b)].
Large curvature of the director field 
further enhances rotation of the director in the island [Fig.\ref{fig:dynamics}(c)],
and two pairs of $\pm 1/2$ defects are created at the periphery of 
the island [Fig.\ref{fig:dynamics}(d)].
The island is broken by creation of defects into two parts of crescent-shape with a pair of defects at their ends.
As they shrink, the defects approach each other 
and annihilate by collision [Fig.\ref{fig:dynamics}(e)(f)].
\Add{To illustrate the mechanism of defect creation, 
we show schematic pictures of the lanes and director field 
in Fig.\ref{fig:dynamics}(g)(h)(i). 
An initially straight lane undergoes a buckling instability 
due to the active stress along the lane boundary, 
forming oppositely tilted regions shown as A and B in Fig.\ref{fig:dynamics}(g).
The director in A and B are tilted clockwise and anticlockwise, respectively.
On the other hand, the active flow at the lane boundaries (arrows) 
generates a clockwise hydrodynamic torque at the center of the lane,
and enhances (suppresses) the tilt in the region A (B).
It causes a rotation of the director in the center of A,
forming the patterns in Fig.\ref{fig:dynamics}(b)(h).
The deformation generates active flow shown by the thick arrow 
in Fig.\ref{fig:dynamics}(h) and further rotates the director field.
as in Fig.\ref{fig:dynamics}(i), where bend deformation is found in 
the encircled regions in the top and bottom (solid lines),
while splay deformation is found on the left and right hand side (dotted lines).
Then, following the scenario of defect creation~\cite{thampi2014instabilities}, 
bend deformation is transformed to splay deformation by 
creating a pair of defects at each of the top and bottom regions.
}

\section{\label{sec::discuss} Discussion}

We studied the effects of an electric field on the 
flow and defect patterns of active nematics
in two dimensions.
The anisotropic active turbulence under \Add{an} electric field is characterized
by flow anisotropy. The magnitude and \Add{degree of} anisotropy of 
\Add{the} flow are both 
maximized at an intermediate field strength.
The decrease of defects with increased field strength 
is in agreement with the previous numerical result 
for three-dimensional active nematics~\cite{krajnik2020spectral}.
We find that 
vortices are localized in the anisotropic active turbulence,
which is reflected in the long tail in the probability distribution of vorticity.
As each localized vortex is isolated, 
topological defects are created and annihilated 
in each vortex region.
This is in contrast with the isotropic active turbulence state where defects 
are created at walls and move along them~\cite{thampi2014instabilities}.

\Add{
A strong electric field stabilizes the uniform aligned state, 
and we determined the threshold by a linear stability analysis.
The most unstable wavevector is parallel to the unperturbed director
along the electric field, which corresponds to a bend deformation.
This is interpreted as a buckling of the director or the filament
due to the extensile active stress~\cite{simha2002hydrodynamic,baskaran2009statistical,vliegenthart2020filamentous}. 
In the three-dimensional numerical simulation~\cite{krajnik2020spectral},
the electric field induced a direct transition between 
active turbulence and uniformly aligned states.
The transition threshold 
was estimated in the unit of 
$E_0 = \sqrt{K/(\varepsilon_a \xi^2)}$ as $E_c/E_0 \simeq 0.7$
for the activity strength $\alpha=0.13 (K/\xi^2)$,
where the notations are translated to those of our model
and $\xi$ is the correlation length for a passive nematic.
Using the expression $\xi = \sqrt{K/|A|}$ for our two-dimensional model, 
the threshold field strength obtained by the linear stability analysis reads 
$E_c/E_0 \simeq 0.19$
for the same activity strength $\alpha=0.13 (K/\xi^2) \simeq 0.0195$.
%
%
%
The threshold is lower than in the previous study,
which might be explained by the difference in the flow-aligning parameter $\lambda$.
We used $\lambda=0.1$, which locates our system deep in the flow-tumbling regime,  
while the value $\lambda=1$ assumed in Ref.~\cite{krajnik2020spectral} 
corresponds to the threshold between the flow-aligning and tumbling regimes~\cite{de1993physics}.
The smaller value of $\lambda$ in our study means 
that the active flow induced by a director undulation has less positive feedback
to amplify the deformation, and that it is suppressed by a weaker electric field. 
On the other hand, the space dimension would not affect 
the stability threshold, because the bend deformation takes place 
in a plane containing the director and thus is essentially two-dimensional.
}

\Add{
A more significant difference from the previous study~\cite{krajnik2020spectral}
is the emergence of the laning state in two dimensions, 
which we found between the anisotropic turbulence and uniformly aligned state.
We interpret it as the result of confinement of the director in a plane.
The absence of lanes in the three-dimensional case suggests that
a sinusoidal director undulation 
might be unstable for out-of-plane deformations,
and the secondary instability could set in 
at the same threshold as the first (linear) instability. 
A weakly nonlinear stability analysis of the laning state in three dimensions
would be useful in verifying this picture.
We also find a temporal oscillation between the active turbulence and laning states.
A lane undergoes a buckling instability due to the extensile active stress along 
the lane boundary. The resultant hydrodynamic torque explains 
local rotation of the director field and simultaneous creation of two pairs of defects. 
}

\Add{
Experimentally, confinement of the active nematics in a quasi-two channel
causes friction between the fluid and walls.
Friction also induces transitions from anisotropic turbulence to a laning state,
as shown both experimentally~\cite{guillamat2016control} and 
numerically~\cite{thampi2014active,srivastava2016negative,thijssen2020active}.
In the experiment~\cite{guillamat2016control}, 
anisotropic friction was realized by a contacting smectic liquid crystal, 
and a periodic switching between a laning state and a disordered flow pattern  
was observed.
The oscillatory behavior resembles our finding, but it should be noted that 
the laning state in the experiment contains arrays of defects and 
is not equivalent  to the defectless laning state found in our simulation.
In the numerical simulations, laning states were induced by
isotropic~\cite{thampi2014active,srivastava2016negative} 
or anisotropic~\cite{thijssen2020active} friction with a substrate.
For isotropic friction, the lane orientation is selected by 
spontaneous symmetry breaking in contrast to the present model.
For anisotropic friction, 
a non-monotonic dependence of the flow anisotropy 
on the strength of friction and temporal fluctuations of the lane speed 
are reported~\cite{thijssen2020active}, which are qualitatively similar to our findings.
However, it should be stressed that friction in 
the previous works did not produce the uniformly aligned state.
We showed for the first time that active nematics can exhibit 
the three states (active turbulence, laning and uniform states)
by tuning a single parameter.
If we simultaneously impose anisotropic friction and an electric field 
in different orientations,
their competition may give rise to more complex patterns,
which we leave for future study.
}

The present study assumed constant and static dielectric anisotropy.
\Add{
A microtubule (MT) is a highly negatively charged biopolymer, and 
has both a permanent dipole moment due to that of tubulin dimers
and an induced dipole moment due to motion of couterions along 
its long axis~\cite{stracke2002analysis,kalra2020all}.
As a result,  MTs are aligned by both AC~\cite{minoura2006dielectric}
and DC electric field~\cite{kim2007active,ramalho2007microtubule}.
In an AC electric field, MTs obtain high conductivity due to the couterion motion, 
which plays a key role in determining their polarization~\cite{minoura2006dielectric}.
Electroorientation in a DC electric field is observed for MTs
adsorbed onto a substrate~\cite{ramalho2007microtubule}
and in kinesin gliding assays~\cite{van2006molecular,kim2007active,isozaki2015control}.
Alignment was achieved by field strengths $E \le 50$ kV/m in the kinesin gliding assays,
while the speed of microtubule was unchanged up to $110$ kV/m~\cite{van2006molecular}
or had no anisotropy due to the field~\cite{dujovne2008velocity},
suggesting that steps in the kinesin ATP cycle are not affected by the electric field~\cite{dujovne2008velocity}.
However, surface-coated kinesins hinder the alignment of MTs 
depending on their density~\cite{kim2007active}.
Motion of MTs in a suspension is also affected by counterion convection~\cite{minoura2006dielectric}, 
electro-osmotic, and electro-thermal flow~\cite{uppalapati2008microtubule}.
A recent experimental study on the response of a dense network of MTs 
reported accumulation of MT bundles under a pulse electric field~\cite{havelka2022lab}.
}
A microscopic model of an active suspension of microtubules and motors would be useful 
in elucidating the dynamic effects of an electric field.
\Add{
In an AC electric field, the intrinsic molecular polarity of MTs
is irrelevant in determining the electric polarity and not sorted by the electric field.
Therefore, it would be legitimate to use the nematic order parameter in describing the \Add{alignment}.
Also, the timescale of electroorientation of individual MTs \Add{(}$\sim 0.1$ s\Add{)}
is much larger than the alternating period of induced electric dipoles~\cite{minoura2006dielectric},
while it is smaller than the characteristic timescale of active vortices ($\sim 10$ s~\cite{sanchez2012spontaneous}).
\Add{Therefore, the present model could be straightforwardly generalized to a model of} 
MT-kinesin suspensions under an AC electric field on a slow timescale, although 
experimental knowledge on filament-motor interaction under an AC field is lacking.
On the other hand, in a DC field or \Add{a} low-frequency AC field, 
the electric polarity of MTs is determined by the intrinsic polarity and 
their electrophoretic motion could dominate active flow, 
which is presumably weakened due to field-induced polarity sorting.
}
The combined effects of 
confinement~\cite{edwards2009spontaneous,shendruk2017dancing}
and an electric field on active turbulence are 
also an interesting topic left for future work. 

\begin{acknowledgments}
Yutaka KINOSHITA acknowledges support from GP-MS at Tohoku University.
\end{acknowledgments}

\appendix

\Add{
\section{\label{sec:eigenvalues} Details of the linear stability analysis}

The eigenvalues 
of the matrix ${\cal M}$ in Eq.(\ref{eq:matrixM})  are given by
\begin{align}
-\mathrm{i} \omega_{\pm}^{\bs{k}}
&=
\frac12
\Biggl[
- 
\left(\frac{k^2}{\mathrm{Re}}  + \frac{D}{\gamma S_0}   \right)
\nonumber\\
&\pm
\sqrt{
\left(\frac{k^2}{\mathrm{Re}}  - \frac{D}{\gamma S_0}   \right)^2
+ 2k^2 S_0 (\lambda c + 1) 
\left[ -
(\lambda c + 1) D + \alpha c 
 \right]
}
\Biggr].
\tag{A1}
\label{eq:eigenvalues}
\end{align}
where we introduced the abbreviation $c= \cos 2\phi$.
The linear growth rate 
$\mathrm{Im}\, \omega_{+}^{\bs{k}}$
is positive if and only if 
\begin{equation}
\alpha c (\lambda c + 1)
>
\left[
 (\lambda c + 1)^2
+ \frac{2}{\gamma S_0^2 \mathrm{Re}} 
\right]
\left(
\varepsilon_a E^2 + 2 KS_0 k^2
\right). 
\tag{A2}
\label{eq:instabilitycondition2}
\end{equation}
where we recalled $D= \varepsilon_a E^2 + 2 KS_0 k^2$.
It is easily seen that the condition is most easily 
satisfied in the long-wavelength limit ($k\to 0$).
To examine the $\phi$-dependence, 
we 
rewrite (\ref{eq:instabilitycondition2})
as 
\begin{align}
\alpha f(c) - D g(c) &> 0,
\tag{A3}
\label{eq:instabilitycondition3}
\\
f(c) &= c (\lambda c + 1), 
\tag{A4}
\\
g(c) &= (\lambda c + 1)^2 + B,
\tag{A5}
\\
B &=\frac{2}{\gamma S_0^2 \mathrm{Re}}.
\tag{A6}
\end{align}
Since $\alpha>0$ by assumption and $D > 0$, $g(c)>0$,
the condition (\ref{eq:instabilitycondition3}) is satisfied 
if and only if $f(c) > 0$ and 
\begin{equation}
\frac{\alpha}{D} > h(c) \equiv \frac{g(c)}{f(c)}.
\tag{A7}
\end{equation}
Since we also assume $0 < \lambda <1$, the condition $f(c)>0$ is equivalent to $c>0$.
To find the minimum of $h(c)$, 
we use 
\begin{align}
h'(c) &= -\frac{1+B}{c^2} + \frac{B}{(c+1/\lambda)^2}.
\tag{A8}
\end{align}
We see that $h(c)$ is a motonotically decreasing function  
in the range $0<c\le 1$.
Therefore, the stability threshold $\alpha_c$ is determined by
substituting $k=0$ and $\phi=0$ into 
 (\ref{eq:instabilitycondition2}), which gives Eq.(\ref{eq:alphac}).
 }

\Add{
\section{\label{sec:distribution} Angle distribution for sinusoidal director undulation}

Here we compute the distribution function $f(\theta)$ of the director angle
for the sinusoidal profile $\theta(x) = \theta_0 \sin kx$.
Without loss of generality,   
we consider the distribution in the range $-\pi/(2k) \le x \le \pi/(2k)$,
where $\theta(x)$ is monotonically increasing with $-\theta_0 \le \theta \le \theta_0$ and 
there is one-to-one correspondence between $\theta$ and $x$.
The probability to find the angle in the infinitesimal range 
$[\theta, \theta + d\theta]$ is given by $f(\theta) d\theta = dx/(\pi/k)$,
where $dx = d\theta/\theta'(x)$ is the corresponding range of the $x$-coordinate. 
Therefore, we obtain
\begin{equation}
f(\theta) 
= \frac{k}{\pi} \frac{dx}{d\theta} 
= \frac{1}{\pi \theta_0 \cos kx} 
= \frac{1}{\pi \sqrt{\theta_0^2 - \theta^2}}
\tag{B1}
\end{equation}
for $|\theta| \le \theta_0$. Note that $f(\theta)$ is trivially zero for $|\theta| > \theta_0$.
}
\bibliography{kinoshita-paper1.bib} 


\onecolumngrid

\twocolumngrid

\onecolumngrid

\twocolumngrid

\end{document}